\documentclass{iaus260}
\usepackage{graphicx}
\pubyear{2009}
\volume{260}  
\pagerange{10--17}
\jname{The R\^ole of Astronomy in Society and Culture}
\editors{D. Valls-Gabaud \& A. Boksenberg, eds.}

\title[Public Perception of Astronomers]                   
{Public Perception of Astronomers: \\ Revered, Reviled and Ridiculed}  

\author[Michael J. West]           
{Michael J. West}        

\affiliation{European Southern Observatory \\ 
                 Alosnso de C\'ordova 3107 \\ 
                 Santiago, Chile \\
                 email: {\tt mwest@eso.org}}
                 
\begin{document}
\maketitle

\begin{abstract}
Society's view of astronomers has changed over time and from culture to culture.  This review discusses some of the many ways that astronomers have been perceived by their societies and suggests ways that astronomers can influence public perception of ourselves and our profession in the future.
\end{abstract}

\firstsection 
\section{Why does public perception of astronomers matter? }

``The greatest present need of astronomy, the grandest science, is not more big telescopes and big observatories, but a more favorable public opinion." 
This sentiment, from a letter written in 1916 to {\it Popular Astronomy} magazine, is as relevant today as it was nearly a century ago.

Our ancestors observed the heavens for religious and secular reasons, with astronomy and astrology largely indistinguishable.  Observations of stars, planets and constellations were used to create calendars that provided agricultural societies with valuable information for the seasonal planting and harvesting of crops, as well as to predict future events or to discern divine messages from the cosmos.   Given the importance of such activities to the cultural identity and even physical survival of ancient people, it is not surprising that sky watchers had prominent roles in their societies.  An abundance of archaeological evidence from around the world attests to the importance of the ancient astronomer.  

In modern times, however, public perception of astronomers began to change as astronomy evolved from an applied to a pure science.  With less prominent roles in their societies, astronomers were forced to seek new forms of financial support for their scholarly activities from governments or wealthy benefactors and to justify their continued value to their fellow citizens.   Although we live today in a time of remarkable astronomical discoveries, as many politicians and businesses know the public's collective memory can be short, and hence astronomers cannot afford to be complacent about our public image.   Public opinion of astronomers is important for many reasons, chief among them:

\begin{enumerate}

\item{} Astronomy is funded by taxpayers or private donors and supported by politicians.  Without their support, many of the great observatories and university astronomy programs around the world would not exist.   But such support is tenuous $-$ per capita spending on astronomy even in the most  developed countries is usually equivalent to the cost of just  one or two cups of coffee per resident $-$ and during periods of economic hardship astronomy can be a tempting target for cost cutting.  Moreover, a significant fraction of our fellow citizens consider astronomy to be a luxury of affluent societies rather than something vital for the human spirit.   For example, during the 2008 presidential campaign in the United States, candidate John McCain criticized Chicago's Adler Planetarium (the oldest in the Western Hemisphere) for requesting government funds to replace its forty-year-old planetarium projector with a new one, a request that McCain labeled as ``foolish."   To ensure continued support, astronomers must strive continuously to make a favorable public impression.    

\item{} Society's perception of astronomers is strongly influenced by the arts, literature, movies and television.  Negative portrayals or stereotypes of astronomers can discourage young people from pursuing careers in the field.   This is an especially important issue in the quest to increase participation by women and other underrepresented  groups in astronomy.  Changing this will require, along with other efforts, transforming the public's understanding of who astronomers are and what they do.

\item{} Astronomers' ability to educate and inspire the public with new discoveries is affected by the way they are viewed as social creatures.  As every successful teacher knows, it is easier to get your message across if your audience likes you, respects you and can relate to you.  Cultivating a positive image of astronomers is therefore a key component of public outreach efforts.  We must also be careful not to confuse the public's interest in astronomy with the public's view of astronomers.  After all, many people like automobiles but do not necessarily have favorable opinions about car salespersons.   

\end{enumerate}

The following sections give an overview of public perception of astronomers in different places and times.  The examples given below are by no means a complete list of the varied and sometimes complex ways that astronomers have been viewed by their societies, and readers should bear in mind that they reflect an unfortunate but natural bias on the author's part towards English-language sources.  Nevertheless, they provide useful insights into the ways that astronomers have been revered, reviled and ridiculed around the world.

\section{The Revered Astronomer}

Astronomers have often enjoyed favorable public opinion and esteemed roles in their societies.  A few examples:

\begin{itemize}
\item{}  ``The ancient Hawaiians were astronomers'' wrote Hawaii's last reigning monarch, Queen Lili'uokalani, in 1897.   For many centuries , the {\it kilo hoku} or `star watchers' were among the most respected members of Hawaii's ruling class.  Crops were planted, fish were caught, wars were fought, and religious festivals were celebrated according the seasonal positions of the stars in the sky or the phases of the moon.   Indeed, it was the {\it kilo hoku}'s specialized knowledge of the positions and apparent motions of the stars that allowed the ocean-faring people of Polynesia to navigate the vast Pacific Ocean and brought them to the Hawaiian islands a millennium ago (see, e.g.,\cite[West 2005]{West 2005}).  As a noted archaeoastronomy expert put it: ``a navigator of the mid-Pacific islands of Oceania would have been as respectable a profession as a neurosurgeon, expert trial lawyer, or perhaps a corporate executive in our culture today"
\cite[(Aveni 1993)]{Aveni 1993}.

\item{} An imperial edict issued in 840 C.E. during the Tang dynasty in China leaves no doubt about the elite status of Chinese astronomers at that time, mandating that ``astronomical officials are on no account to mix with civil servants and common people in general" (\cite[Needham 1959]{Needham1959}).

\item{} In 2005, the {\it New York Times} newspaper surveyed Americans to identify the most prestigious occupations in the United States.   Not surprisingly, doctors and lawyers topped the list of 447 different occupations.  However, the fifth most prestigious occupation was ``astronomers and physicists," which ranked higher than biologists (13), psychologists (19), mathematicians (48) and all other sciences.

\item{}In 2000, {\it Time} magazine chose Albert Einstein as the ``Person of the Century," beating out many other 20th century luminaries.   {\it Time} chose the famed astrophysicist because ``He was the embodiment of pure intellect, the bumbling professor with the German accent, a comic cliche in a thousand films. Instantly recognizable, like Charlie Chaplin's Little Tramp, Albert Einstein's shaggy-haired visage was as familiar to ordinary people as to the matrons who fluttered about him in salons from Berlin to Hollywood. Yet he was unfathomably profound $-$ the genius among geniuses who discovered, merely by thinking about it, that the universe was not as it seemed."   Other astronomers have also appeared on the cover of {\it Time} magazine over the years, including Harlow Shapley (1935), Edwin Hubble (1949), Maarten Schmidt (1966), and Carl Sagan (1980).  

\item{}  Some astronomers have had great success as popularizers of astronomy, attracting wide audiences and public acclaim thanks to their ability to communicate in an informative and entertaining way.  One of the most prolific was the 19th century astronomer Richard Proctor, who wrote more than 60 books during his lifetime and gave thousands of public lectures around the world.  When Proctor died in 1888, an obituary described him as a man ``whose name as an expositor of science has become a household word wherever the English language is spoken" and another said that he had ``probably done more than any other man during the present century to promote an interest among the ordinary public in scientific subjects."  

\qquad Other great astronomy popularizers of the past two centuries have included Camille Flammarion, Agnes Mary Clerke, Arthur Eddington, Percival Lowell, Harlow Shapley, Patrick Moore, Carl Sagan, Hubert Reeves, Stephen Hawking, Martin Rees, Neil deGrasse Tyson, and others.  Flammarion's book, {\it L'astronomie Populaire}, first published in 1879, sold 100,000 copies, a phenomenal number for that era.  Hawking's 1988 book {\it A Brief History of Time} has sold more than 10,000,000 copies to date, reaching a far vaster audience than Flammarion could ever have imagined.  Like Einstein, Hawking has become an icon of popular culture today, appearing, for example, in episodes of {\it The Simpsons}, {\it Star Trek: The Next Generation} and other television shows.  

\item{} Carl Sagan's {\it Cosmos} was one of the most successful science programs in television history, seen by more than 500 million people worldwide.  Sagan's enthusiasm, charisma and knack for clear explanations of complex scientific ideas were major reasons for the show's success.  A companion book based on the television series remained on best-seller lists for over a year.  Dubbed ``the Cosmic Explainer" by {\it Time} magazine in 1980, Sagan became astronomy's most recognized face, aided by his 25 television appearances on the {\it Tonight Show with Johnny Carson} that reached an audience of millions.  

\begin{figure}[b]
\begin{center}
\includegraphics[width=4.5in]{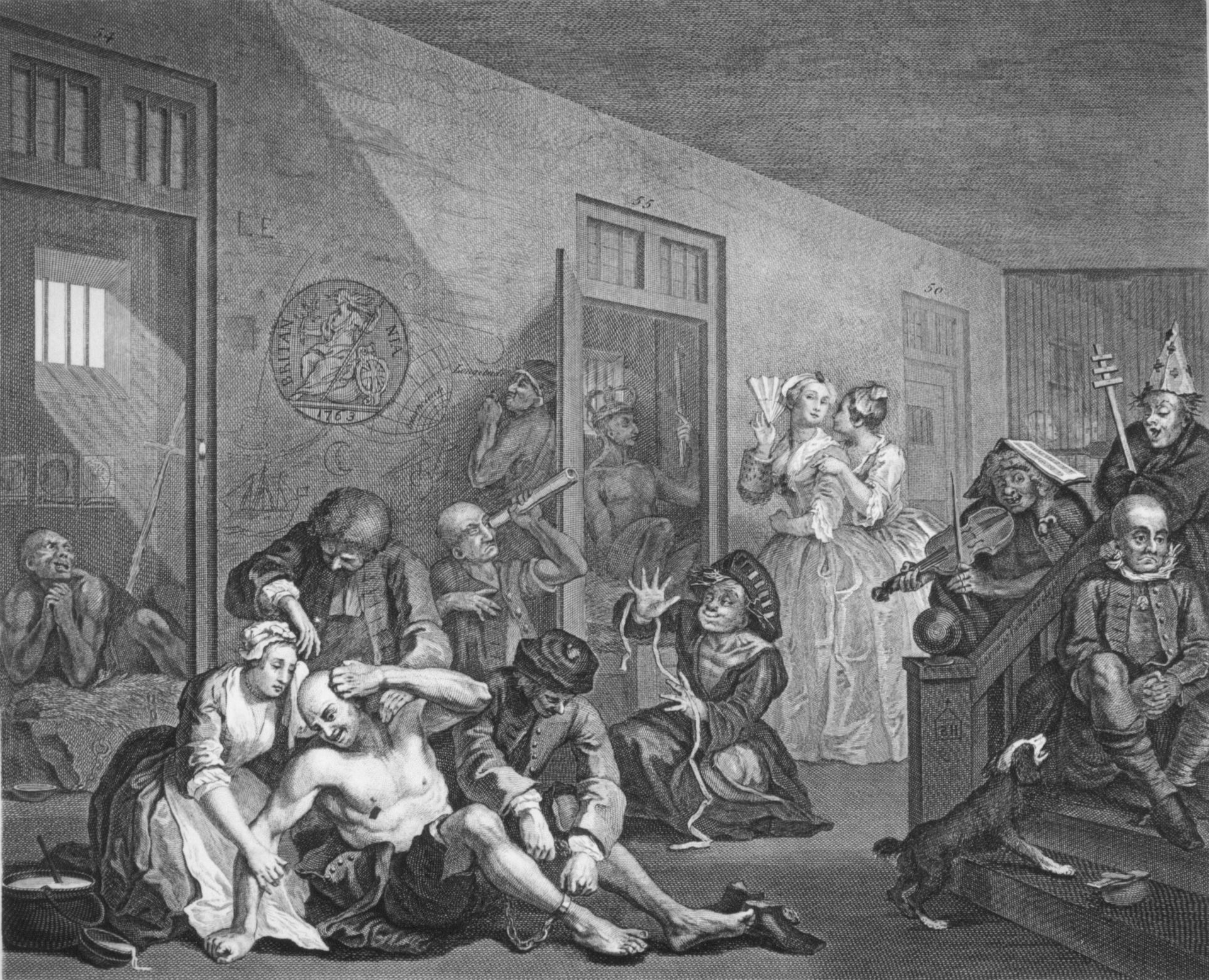}
 \caption{An engraving titled ``The Interior of Bedlam," from {\it A Rake's Progress} by William Hogarth, 1735.  In the center of the image, a mad astronomer looking through his telescope is one of the delusional inhabitants of an institution for the mentally ill.  In the public perception of mid-18th century Britain, astronomy could drive even the sanest minds insane.  A similar view was expressed two centuries later by the American journalist and social commentator H.L. Mencken, who wrote ``in the popular belief, most astronomers end by losing their minds"
\cite[(Mencken 1956)]{Mencken 1956}.}
\label{fig1}
\end{center}
\end{figure}

\end{itemize}

\section{The Reviled Astronomer}

Although astronomers have enjoyed a fairly benign public image in general, they have occasionally been the subject of societal scorn or wrath.  Some examples:

\begin{itemize}

\item{} On October 4, 1876, the {\it New York Times} published a scathing editorial titled {\tt The Fall of an Astronomer.}  In it, the {\it Times}  harshly criticized renowned French astronomer Urbain Le Verrier for falsely promising that he would provide proof of his claimed discovery of the hypothetical planet Vulcan, complaining that ``he has broken his word, and has trifled with our holiest planetary emotions."   The editorial concluded with a stinging condemnation: ``there is, perhaps, no spectacle more sad than that of the once honored astronomer who has lost the confidence of his fellow-men and is sneered at with impunity even by the sidewalk scientific person who shows the moon through a cracked telescope to astonished rural visitors."

\item{}The scientific method has, of course, had its detractors and astronomers have not been immune to occasional criticism or even vilification.  For example, American poet Walt Whitman's famous 1865 poem {\it When I Heard the Learn'd Astronomer} describes his experience attending a public astronomy lecture and how he became ``tired and sick" from the endless proofs,  charts, diagrams and measurements presented by the astronomer until he finally escaped to the ``mystical moist night air" and ``looked up in perfect silence at the stars."   Although he was not anti-science, Whitman, like  other writers and artists during the 19th century English Romantic and American Transcendentalist movements, believed that the deepest truths are revealed through feelings and spiritual experience and that science devalues the innate beauty of nature.  {\it When I Heard the Learn'd Astronomer} was recently brought to a new and younger audience in the English-speaking world with the publication in 2004 of an award-winning illustrated children's book based on Whitman's poem (Whitman \& Long 2004), with color pictures showing a young boy whose passion for astronomy turns to boredom when he attends an astronomy lecture.

\item{} When Halley's Comet returned in 1910, its passage near Earth led to sensationalist reports in the popular media of impending doom from cyanogen gas in the comet's tail or a direct collision with our planet.   Public suspicion that astronomers were hiding the truth about the dangers led to frustration and accusations.  For example, an article published in the {\it New York Times} on April 24, 1910 was titled {\tt Astronomers Suspected} and discussed the flood of letters received by Harvard Observatory.  The exasperation felt by astronomers is evident from E.C. Pickering, the observatory's director, who said ``they seem to think we are responsible for it.  Insinuations in several letters we have received made us fear we were being denounced as undesirable citizens." 

\item{} Construction of astronomical observatories on mountaintops considered sacred by indigenous people or in environmentally sensitive locations has sometimes been a source of controversy, putting astronomers in an unfavorable light and open to public criticism.  An example is Mauna Kea,  home to some of the world's greatest telescopes but also a sacred site to some native Hawaiians.  A small but vocal group of Hawaiians and environmentalists has long opposed the presence of observatories atop Mauna Kea, and this opposition has included protests and legal challenges.   A 2008 public opinion poll found that a strong majority of Hawaii residents favor building new larger telescopes atop Mauna Kea, but only if it is done in a way that is sensitive to the concerns of Hawaii's indigenous culture.  Similarly, the creation of Mount Graham Observatory in 1988 led to controversy with the Apache people of the southwestern United States who consider the mountain a holy site, and with  environmentalists concerned about the impact of observatory construction on an endangered species of squirrel.  Although such conflicts are, unfortunately, sometimes unavoidable when differing world views collide, anytime that astronomers are perceived $-$ rightly or wrongly $-$ as thwarting the rights of indigenous people or protectors of the environment there is a risk of backlash in the court of public opinion.  Today Mauna Kea and Mount Graham are home to some of the world's most sophisticated telescopes but also to lingering resentment from some members of their local communities.

\item{} Astronomers faced a public relations disaster in 2006 following the International Astronomical Union's decision to strip Pluto of its status as the solar system's ninth planet.  Public anger was palpable, fueled by the loss of a cherished icon and the obvious bickering among astronomers.  Letters to newspapers, phone calls to radio shows, and internet blogs criticized astronomers for being mean-spirited and capricious.  Just a few years earlier, the public had rallied around astronomers when the Hubble Space Telescope was threatened by NASA's decision to cancel future servicing missions to the orbiting observatory, and the public outcry was surely an important factor in the eventual decision to save HST.  But whereas the public could blame NASA administrators for the initial decision to abandon HST, it was astronomers themselves who were responsible for Pluto's demotion and consequently there was little goodwill or understanding from the public.  

\qquad A review of the Pluto fiasco by astronomers David Jewitt and Jane Luu concluded: ``Public perception of the process, and of astronomers and astronomy, has been soiled.  Millions of people now think of astronomers as having too much time on their hands, and as unable to articulate the most basic definitions... None of this is good for astronomy" 
\cite[(Jewitt \& Luu 2007)]{JL2007}.

\end{itemize}

\section{The Ridiculed Astronomer}

Perhaps the most common view of astronomers has been one of amusement, often based on a misunderstanding of what astronomers do or recurring stereotypes of scientists in general.  
 A 2002 report by the U.S. National Science Board concluded ``It is generally conceded that scientists and engineers have somewhat of an image problem.  Although their intelligence and work are highly respected, that admiration does not seem to extend to other aspects of their lives. The charming and charismatic scientist is not an image that populates popular culture" and noted that scientists are almost always portrayed by the entertainment industry  as ``unattractive, reclusive, socially inept white men or foreigners working in dull, unglamorous careers."  
Some examples from the past and present:

\begin{itemize}
\item{} The world's oldest surviving book of jokes, {\it Philogelos} (``The Laughter-Lover"), which dates from around the fourth century C.E., contains several jokes about astrologers and their misreadings of the heavens.   In a similar spirit, an Aesopic fable from ancient Greece mocks an absent-minded astrologer who falls into a hole one night while gazing at the stars.  A passerby who comes to his aid chastises him: ``While you are trying to see what is in the heavens, don't you see what is on the earth?" 

\item{} {\it Gulliver's Travels}, Jonathan Swift's 18th century satirical novel, ridicules the astronomers 
of the fictitious floating island of La Puta as ``never enjoying a minute's peace of mind" because they worry obsessively about possible changes to celestial bodies.  So great are the astronomers' fears that ``they can neither sleep quietly in their beds, nor have any relish for the common pleasures or amusements of life. When they meet an acquaintance in the morning, the first question is about the sun's health, how he looked at his setting and rising, and what hopes they have to avoid the stroke of the approaching comet.''  Gulliver also remarks that most of the La Putan astronomers believe in astrology but are too ashamed to admit it, and they have an unfortunate tendency to be outspoken about politics and other topics in which they have no expertise. 

\item{}Antoine de Saint-Exup\'ery's classic children's book {\it Le Petit Prince} describes the adventures of the Little Prince as he journeys to Earth from his tiny asteroid home around a distant star.  This asteroid, readers are told, has been seen only once through a telescope  by a Turkish astronomer in 1909.  Unfortunately, nobody believed the astronomer when he presented his discovery to the International Astronomical Congress because he dressed in Turkish clothes.  Only when he presented the same discovery years later, this time dressed in European clothes, was the Turkish astronomer believed.  ``Grown-ups are like that,"  readers are told.  Since its publication in 1943, {\it Le Petit Prince} has been translated into nearly 200 languages and sold more than 80 million copies worldwide, influencing the way countless children have perceived silly astronomers.

\item{} For the past century, movies and television have provided an especially powerful visual medium that both reflects and shapes public perception of astronomers.   
One of the pioneers of early cinema, Georges M\`eli\'es made two seminal films with astronomers as the main characters.  His 1898 film {\it La Lune \`a Un M\`etre} (known as {\it The Astronomer's Dream} in English) tells the story of an astronomer's strange dream after he falls asleep in his observatory.   In his dream, the Moon taunts the astronomer with a wide grin and mischievous eyes before eating his telescope and eventually devouring the astronomer too.  M\`eli\'es' 1902 masterpiece {\it Le Voyage Dans la Lune (A Trip to the Moon)} tells the story of six astronomers that travel to the moon, where they encounter and fight a few moon creatures before returning safely to Earth.  In both films, all the astronomers are older gentlemen, some with long white beards.  

\qquad A few years later, Gaston Velle's whimsical film {\it Voyage Autour d'une Etoile (Voyage Around a Star)} featured a daffy astronomer who becomes infatuated with the stars and floats up to the heavens inside a giant soap bubble to be with them.  He remains with the stars, which are represented as beautiful women, until an angry celestial god tosses the human back to Earth, where he lands impaled on a rooftop weathervane.  

\begin{figure}[t]
\begin{center}
\includegraphics[width=5.0in]{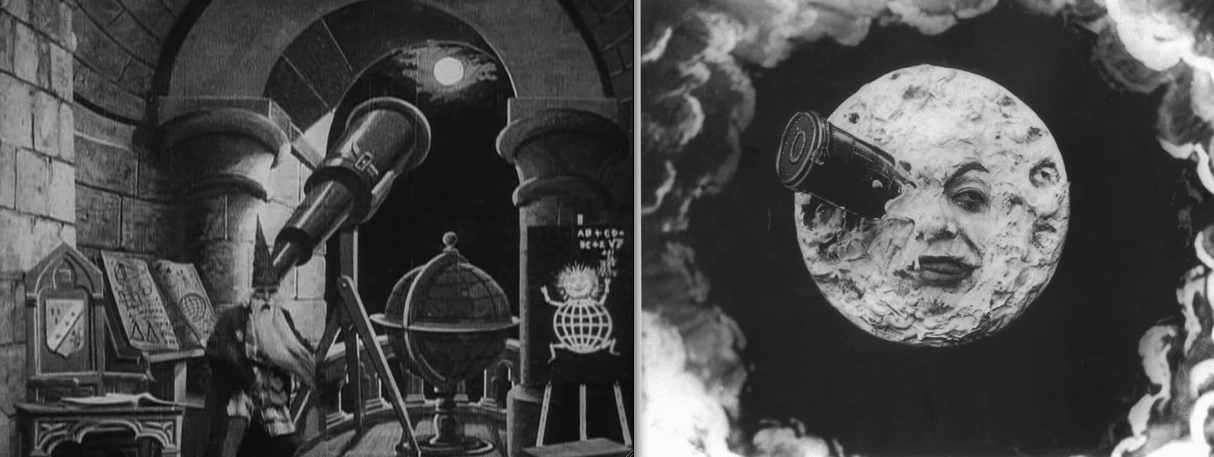}
 \caption{Scenes from two of Georges M\`eli\'es' seminal movies, {\it La Lune \`a Un M\`etre (The Astronomer's Dream)} from 1898 and {\it Voyage dans la Lune (A Trip to the Moon)} from 1902.
 Both films featured astronomers as their main characters. }
\label{fig2}
\end{center}
\end{figure}

\qquad On the lighter side, 
in the romantic comedy {\it The Heavenly Body} (1944), actor William Powell played an absent-minded astronomer who is so focused on his research that he neglects his beautiful wife played by Hedy Lamarr until, in desperation, she turns to an astrologer for advice.  

\qquad It is noteworthy that, despite the under-representation of women astronomers in many countries,  a few Hollywood films such as  {\it Roxanne} (1987),  {\it Contact} (1997) and the 2008 remake of the classic sci-fi film {\it The Day the Earth Stood Still} have featured women astronomers or astrobiologists as main characters.  

\item{} Scientists are frequently stereotyped as social misfits, an idea perpetuated by the entertainment industry.   Indeed, a survey of Americans found that people who watch more television have a greater tendency to view scientists as odd (Gerbner \& Linson 1999).  

\qquad A recent example of this sort of stereotyping is {\it The Big Bang Theory}, an American television show that premiered in 2007 and which currently averages nine million viewers each week plus a growing international audience.  This situation comedy follows the daily lives of four socially maladjusted science geeks, including a string theorist and an astrophysicist who happen to live next door to an attractive young woman.   
Despite the show's advertising slogan `smart is the new sexy' the portrayal of the scientists is often funny but rarely flattering.  In a 2008 review of the show, the {\it Wall Street Journal} described one of the main scientist characters as having ``a huge IQ but no aptitude for social niceties nor discernible interest in the opposite sex" and the other scientist as ``yearning for social acceptance."  By contrast, their non-scientist neighbor is described as ``at ease in the world, a sucker for hunks and, in her often mortified response to the guys' brainy antics, a proxy for viewers.''   

\item{} In 1908, Lick Observatory Director W.W. Campbell lamented ``The opinion prevails quite generally that an astronomer's duty consists in sitting all night with his eye at the end of the telescope, sweeping the heavens, in order to discover new bodies - comets, planets, moons and new stars; and that this is the end and aim of science. This view is far from the truth."    

\qquad Unfortunately, astronomers have been portrayed this way endlessly in the popular media, leading to some confusion in the public's mind about what astronomers actually do.   To give just a few of myriad examples, the cheesy 1959 science fiction B-movie {\it Teenagers from Outer Space} begins with an astronomer in a major observatory gazing through the eye piece of a telescope, casually searching the sky for interesting new objects until by chance he discovers an alien spaceship headed to Earth.  Similarly, the 1956 movie {\it Uchujin Tokyo Ni Arawaru} (released in English as {\it Warning from Space}) features Japanese astronomers in white laboratory coats who are startled to discover an approaching alien spacecraft while peering through an enormous telescope.
And in {\it Contact}, a movie based on Carl Sagan's novel that has earned more than \$170,000,000 USD in worldwide sales, a radio astronomer detects a message from an extraterrestrial civilization while listening to radio signals from outer space, unintentionally perpetuating a common misconception that radio waves are sound waves.  
\end{itemize}

\section{Influencing Public Perception of Astronomers}

Given the importance of public opinion for the future of astronomy, what can be done to foster a positive and accurate view of our profession?  Here are a few suggestions:

{\it Be the public face of astronomy.}   Most of us were inspired to become astronomers because of something or someone we read, heard or saw when we were younger.  Each of us has the potential to change lives in the same way by inspiring the next generation of astronomers.   Although we can rely on our public outreach colleagues and astronomy-friendly journalists to assist, in the end astronomers must be the public face of astronomy.    
By humanizing astronomy and astronomers $-$ telling our own story $-$ we can connect with the public and in the process help to manage our own image.

{\it Learn to be an effective communicator.}  Most astronomers have had little or no training on how to communicate effectively with the public.  Consequently, many are uncomfortable giving interviews or public lecture, or talk in a way that is difficult for non-scientsts to understand.  Few people are born gifted science communicators, but most people can learn to be proficient at it.  At ESO, for example, we have been providing media training for our astronomers  to help them to become better science communicators.   Obviously some astronomers are more skilled as writers than public speakers, and others are more comfortable working with children than with adults.  Each must decide on his or her own outreach strengths and then develop them accordingly.

{\it Embrace new technologies and opportunities to connect with the public.}  We live in a fast-moving world. Today's cutting-edge technologies such as Twitter, Facebook, and Google will undoubtedly seem dated in the not-too-distant future, replaced by newer and more powerful means of mass communication.  Although books and magazines remain effective tools for presenting astronomy and astronomers to a wide audience, the sad fact is that reading rates have been declining for years in the United States (a 2008 study by the National Endowment for the Arts found that nearly half the American adult population had not read a book during the previous year).  Yet at the same time novels that can be downloaded and read on mobile phones have exploded in popularity in Japan, in large part because they fit current lifestyles.  To get our message to the public, 
astronomers must not only adopt new technologies but also find creative new ways to use them.  A good example is astronomy popularizer Phil Plait's {\it Bad Astronomy} website, which was chosen by {\it Time} magazine as one of the 25 best blogs of 2009, citing him as ``a voice of reason amidst the nonsense of non-science."  

The bottom line is that astronomers must actively define our public image, otherwise it will be defined for us.  Although it is impossible to control how astronomers are perceived by society, we do have some power to influence our public image.

\end{document}